\documentstyle[sprocl]{article}
\begin{document}   
\newcommand{\ul}{\underline}
\newcommand{\nid}{\noindent}
\newcommand{\be}{\begin{equation}} 
\newcommand{\ee}{\end{equation}} 
\newcommand{\ds}{\displaystyle}
\newcommand{\pad}{\partial} 
\def\Journal#1#2#3#4{{#1} {\bf #2}, #3 (#4)}

\def\PRD{{\em Phys. Rev.} D}

\font\eightrm=cmr8

\bibliographystyle{unsrt} 

\pagenumbering{arabic}
\vskip 50pt 

\centerline{\large 
2$^{nd}$ Amaldi Conference on Gravitational Waves, 1-4 July 1997, CERN 
      Geneve } 
\vskip 50pt

\centerline{\nid\LARGE\bf 
On the perturbed Schwarzschild geometry} 
\vskip 4pt
\centerline{\nid\LARGE\bf 
for determination of particle motion
}

\vskip 30pt
\centerline{\Large Alessandro D.A.M. Spallicci}

\vskip 20pt
\centerline{\large Gravitation Research Group, 
Salerno Univ. at Benevento}

\vskip 10pt
\centerline{\large Kamerlingh Onnes Lab., Dept. Phys.\& Astr., 
Fac. Math.\& Nat. Sc., 
Leiden Univ.}

\vskip 25pt
\centerline{\large\bf Abstract} 
{\nid\large A novel method for calculation of the motion and  
radiation reaction for 
the two-body problem (body plus particle, 
the small parameter {\small $m/M$} being the ratio of the masses) is 
presented. 
In the background curvature given by the Schwarzschild 
geometry rippled by gravitational waves, the geodesic equations 
insure the presence of radiation reaction also 
for high velocities and strong field.
The method is generally    
applicable to any orbit, but radial fall is of 
interest due to the non-adiabatic regime (equality of 
radiation reaction and fall time scales), in which the particle 
locally and immediately reacts to the emitted radiation. 
The energy balance hypothesis is only used (emitted radiation equal to the variation in the kinetic 
energy) for determination of the 4-velocity via the Lagrangian and 
normalization of divergencies. 
The solution in time domain of 
the Regge-Wheeler-Zerilli-Moncrief radial wave equation 
determines the metric tensor expressing the polar perturbations, in terms  
of which the geodesic equations are written and shown herein.
}

\pagebreak

\pagenumbering{arabic}

\title{ON THE PERTURBED SCHWARZSCHILD GEOMETRY FOR DETERMINATION OF PARTICLE 
MOTION}

\author{ALESSANDRO D.A.M. SPALLICCI}

\address{Gravitation Research Group, 
Salerno Univ. at Benevento}

\address{Kamerlingh Onnes Lab., Dept. Phys.\& Astr., 
Fac. Math.\& Nat. Sc., 
Leiden Univ.}

\maketitle\abstracts{ 
A novel method for calculation of the motion and  
radiation reaction for 
the two-body problem (body plus particle, 
the small parameter {\small $m/M$} being the ratio of the masses) is 
presented. 
In the background curvature given by the Schwarzschild 
geometry rippled by gravitational waves, the geodesic equations 
insure the presence of radiation reaction also 
for high velocities and strong field.
The method is generally    
applicable to any orbit, but radial fall is of 
interest due to the non-adiabatic regime (equality of 
radiation reaction and fall time scales), in which the particle 
locally and immediately reacts to the emitted radiation. 
The energy balance hypothesis is only used (emitted radiation equal to the variation in the kinetic 
energy) for determination of the 4-velocity via the Lagrangian and 
normalization of divergencies. 
The solution in time domain of 
the Regge-Wheeler-Zerilli-Moncrief radial wave equation 
determines the metric tensor expressing the polar perturbations, in terms  
of which the geodesic equations are written and shown herein.
}

\section{Introduction}

Regge and Wheleer \cite{RW1957} proved the vacuum stability of 
the Schwarzschild black hole, while  
Zerilli \cite{Z1970,Z1975}
studied the  
emitted radiation via the radial wave equation 
for polar perturbations which source   
is a freely 
falling test mass {\small $m$} towards the black hole of large 
mass {\small $M$}. 
Moncrief \cite{M1974} showed the gauge invariant significance of 
the wave equations.
Past work was concerned on the Fourier analysis 
of the emitted radiation (for a 
review see Ruffini \cite{R1972}) and not on the 
motion of the particle. 
Radiation reaction for particles around black holes has been   
tackled by Cutler and co-workers \cite{Cu1994} whose analysis 
(via Teukolsky formula) is based 
on the adiabatic approximation, the postulate of energy-balance,  
and is limited to {\small $6~M$}; 
for the Dirac-Galts'ov method \cite{Ca1997}   
antitrasformation of all frequencies is required \cite{S1998}, 
while the axiomatic approach 
\cite{QW1997}
is conceived within the {\it one-body} problem. 
The aim of this work is the identification of  
radiation reaction, 
without the assumption of adiabacity,
with minimal use of the energy balance 
postulate, and identification of the 
trajectory up to the horizon. Future developments
may include a general method for motion of small objects 
in any orbit;  
the solution of 2nd-order 
equation, which energy-momentum tensor is based on the geodesic equations  
calculated herein, and a post-Schwarzschildian formalism.

\section{Approach}

Past work was confined to a particle falling in   
an {\it unperturbed} Schwarz\-schild metric, but 
radiation reaction requires a geodesic path through   
the emission of radiation, in a {\it perturbed} Schwarzschild metric. To  
this end it appears necessary:   
a numeric or analytic 
solution of Regge-Wheeler-Zerilli-Moncrief (RWZM) equation in time domain; 
determination of the perturbation components and of 
1st-order Lagrangian; finally identification of radiation reaction 
by subtraction of the 0th-order terms in the geodesic equations. 
The RWZM equation for polar perturbations is: 

{\small
\be
\frac {d^{2} \Psi_l(r,t)}{dr^{*2}} -  
\frac {d^{2} \Psi_l(r,t)}{dt^2} - V_{l}(r)\Psi_l (r,t) = S_{l}(r,t)
\label{eq:rwzm*}
\ee
}

where {\small 
$r^{*}= r + 2M \ln \left ({\ds{r\over 2M}} - 1 \right )$} is the 
tortoise~coordinate and the potential 
{\small $V_{l}(r)$} is:

{\small
\be
V_{l}(r) =  \left ( 1- \frac{2M}{r}\right ) 
\frac {2\lambda^{2}(\lambda + 1)r^{3} + 6\lambda^{2}Mr^{2} + 
18\lambda M^{2}r + 18M^{3}} 
{r^{3}(\lambda r + 3M)^2}
\ee
}

Further {\small $\lambda = {1\over 2} (l - 1) (l + 2)$} and 
{\small $S_{l}(r,t)$} is the {\small $2^{l}$}-pole 
source component and, for a radially 
falling particle, is:


{\small
\be
S_{l}\! =  
\frac {\left(1\!-\!{\ds\frac{2M}{r}}\right ) k}{(\lambda\!+\!1)
(\lambda r+\!3M)}
\!\left \{\! r\left(1\!-\!{\ds\frac{2M}{r}}\right ) 
\delta '[r\! - \! r_0(t)] 
+
\!\!\left [
\frac{6M \left(1\!-\!{\ds\frac{2M}{r}}\right)^2}
{\lambda r + 3M} 
+ \! \frac{M}{r} - \lambda\! - \! 1 \right ]\!
\delta [r -r_0 (t)]
\right \}
\ee
}

where {\small $k = 4M\sqrt{(2l + 1)\pi} $} and {\small $r_0(t)$} 
is the inverse of: 

{\small
\be
t = - 4M\left({\frac{r}{2M}}\right )^{1/2} 
- \frac{4M}{3}\left(\frac{r}{2M}\right )^{3/2} - 2 M
\ln \left [\left ( \sqrt{\frac{r}{2M}} - 1 \right ) 
\left (
\sqrt{\frac{r}{2M}} + 1 \right )^{-1} \right ] 
\ee
}
Eq.(\ref{eq:rwzm*}) can be rewritten in terms of {\small$ t,r^*$}  
and with constant coefficients of the 2-nd derivatives but solely    
via an approximate 
inverse function {\small $r(r^*)$} and thus resulting into   
an approximate p.d.e. (further, the solution is  
most interesting at {\small $r^* = \infty$}). In  
the {\small $(t,r)$} domain instead,
eq.(\ref{eq:rwzm*}) becomes: 

{\small 
\be
\frac{1}{A^2(r)}\frac {d^{2} \Psi_l}{dr^2} -  
\frac{1 - A(r)}{rA^2(r)}\frac {d \Psi_l}{dr} -  
\frac {d^{2} \Psi_l}{dt^2} - V_{l}(r)\Psi_l = S_{l}(r,t)
\label{eq:rwzm}
\ee
}
where {\small $A(r) = {\ds\frac{dr^*}{dr}} = {\ds\frac{r}{r - 2 M}}$}. 
The polar perturbations 
{\small $h_{\mu\nu}$} are determined by:

{\small
\be
\left(\matrix{
\left (\!1\!-\!\frac{\displaystyle 2M}{\displaystyle r}\!\right )
\!H_{0} 
& H_{1}
& h_0 \frac{\partial}{\partial \theta} 
& h_0 \frac{\partial}{\partial \varphi} \cr
sym 
& \left (\!1\!-\!{\ds\frac{2M}{r}}\!\right )^{-1}\!H_{2}
& h_1 \frac{\partial}{\partial \theta} 
& h_1 \frac{\partial}{\partial \varphi} \cr
sym
& sym
& r^2 \!\!\left [\! K\! +\! G 
{\ds\frac{\partial ^ 2}{\partial \theta  ^2}} \!
\right ]\!\! 
& sym \cr
sym
& sym
& r^2 G\!\left ( \!
\frac{\displaystyle \partial ^ 2}
{\displaystyle \partial \theta \partial \varphi }\! - \!
\cot\!\theta \frac{\displaystyle \partial} 
{\displaystyle \partial \varphi}\! \right )\! 
& r^2\!\!\left [\! K\! \sin^2\!\theta\! + \!G\!\left ( 
\!\frac{\displaystyle \partial ^ 2}
{\displaystyle \partial \varphi ^2}\! + \!
\sin\!\theta \cos\! \theta \frac{\displaystyle \partial} 
{\displaystyle \partial \theta}\! \right )\! \right ]\!\! 
}\right ) Y_l^z
\ee
}

where {\small $H_0, H_1, H_2, h_0, h_1, K, G$} are functions of 
{\small $t, r$}. The 
Regge-Wheeler gauge specifies {\small $G = h_0 = h_1 = 0$}. There 
are relations between {\small $\Psi$} and derivatives 
and the perturbation components. 
Analytically the initial value problem is well defined 
({\small $\Psi$} and {\small $\dot\Psi$} are zero, i.e. 
particle at rest at infinity). Numerically, Lousto and Price
\cite{LP1997} have determined intermediate conditions, 
solving eq. (\ref{eq:rwzm*}) with a finite difference scheme and 
direct integration of the source term (thus    
an interpolation should be performed for   
finding 
{\small $\Psi$ and $\dot\Psi$}). Alternatively for    
an analytic solution of eq.(\ref{eq:rwzm}), the asymptotic  
solutions for {\small $ r\rightarrow \infty $ and $r\rightarrow 0$} 
would be instrumental for 
qualitative methods as differential inequalities or Lagrange   
identities techniques \cite{FR1996}.

\section{The equations of motion}

The Schwarzschild metric up to  
1st-order perturbations in the Regge-Wheeler gauge is: 

{\small 
\be
ds^2\!=\!ds_0^2 
+ 
\left (\!1\!-\!{2M\over r}\!\right)\!H_0Ydt^2 
+\left (\!1\!-\!{2M\over r}\!\right)\!\!^{-1}\!H_2Ydr^2 
+ r^2\!KY d\theta^{2} 
+ r^2\! \sin^2\!\theta KY d\phi^2 
+ 2 H_1 Ydtdr
\ee
}
The geodesic equations encompass three perturbation schemes: 
i) a perturbative field 
{\small $g^{\mu\nu} = \eta_{Schwarzschild}^\mu\nu + 
h^{\mu\nu}_{1st-order~perturbation}$};
ii) the particle trajectory {\small $r_p(t) 
= r_0(t) + r_1(t) = r_0 [1 + \epsilon _r(t)]$ }
where {\small $r_0(t)$} is the trajectory 
crossing the Schwarzschild metric and radiating out 
but the motion being unperturbed and {\small $r_1(t)$} is the  
correction due to radiation reaction, 
via the small parameter {\small $\epsilon _r (t)$};
iii) the field at {\small $r_p(t)$} is a McLaurin series at  
{\small $r_0,t_0$}, and 
{\small $t_p = t_0 + t_1 = t_0 [1 + \epsilon _t(t)]$}:
{\small
$
g^{\mu\nu}(r_p, t_p) \simeq
= 
g^{\mu\nu}(\!r_0,t_0\!) 
+ \epsilon_r r_0  \frac{\pad g^{\mu\nu}}{\pad r}\!\!\mid _{(\!r_0,t_0\!)} 
+ \epsilon_t t_0  \frac{\pad g^{\mu\nu}}{\pad t}\!\!\mid _{(\!r_0,t_0\!)} 
$
}

There are two unknown variables {\small $\epsilon_r(t)$} and   
{\small $\epsilon_t(t)$}. For the case of radial fall 
$d\theta$ and $d\phi$ vanish, and   
using the time independence of the Schwarzschild metric, 
and dropping the $(r_0,t_0)$ notation
one equation is:

{\small
\[
\frac{d^2 [(1 + \epsilon_r)r_0]}{ds^2} 
+ {1 \over 2} 
\left ( 
\eta^{rr}  
+ \epsilon_r r_0 {\ds\frac{\pad \eta^{rr}}{\pad r}} 
+ h^{rr} 
+ \epsilon_r r_0 {\ds\frac{\pad h^{rr}}{\pad r}} 
+ \epsilon_t t_0 {\ds\frac{\pad h^{rr}}{\pad t}}  
\right ) \times 
\]
\[       
\left \{ \pad\left [
{
{ 3 \left (
\eta_{r  r } 
+ \epsilon_r r_0 {\ds\frac{\pad \eta_{r  r }}{\pad r}} 
+ h_{r  r } 
+ \epsilon_r r_0 {\ds\frac{\pad h_{r  r }}{\pad r}} 
+ \epsilon_t t_0 {\ds\frac{\pad h_{r  r }}{\pad t}}  
\right )} 
\over {\pad r }} 
\right ]  
\frac{dr}{ds}\frac{dr }{ds}  
\right .
\]
\[
+ \pad \left [
{ 
{\left (
h_{rr} 
+ \epsilon_r r_0 {\ds\frac{\pad h_{rr}}{\pad r}} 
+ \epsilon_t t_0 {\ds\frac{\pad h_{rr}}{\pad t}}  
\right )} 
\over {\pad t}} 
+ {
{ 2 \left (
h_{rt} 
+ \epsilon_r r_0 {\ds\frac{\pad h_{rt}}{\pad r}} 
+ \epsilon_t t_0 {\ds\frac{\pad h_{rt}}{\pad t}}  
\right )} 
\over {\pad r}} 
\right ] \frac{dr}{ds}\frac{dt}{ds}  
\]
\[
+ \pad 
\left [
+ { {2 \left (
h_{rt} 
+ \epsilon_r r_0 {\ds\frac{\pad h_{rt}}{\pad r}} 
+ \epsilon_t t_0 {\ds\frac{\pad h_{rt}}{\pad t}}  
\right )} 
\over {\pad r}} 
+ {{\left (
h_{rr} 
+ \epsilon_r r_0 {\ds\frac{\pad h_{rr}}{\pad r}} 
+ \epsilon_t t_0 {\ds\frac{\pad h_{rr}}{\pad t}}  
\right )} 
\over {\pad t}} 
\right ]
\frac{dr}{ds}\frac{dr}{ds}  
\]
\[
\left .
+ \pad \left [
{{2 \left (
h_{rt} 
+ \epsilon_r r_0 {\ds\frac{\pad h_{rt}}{\pad r}} 
+ \epsilon_t t_0 {\ds\frac{\pad h_{rt}}{\pad t}}  
\right )} 
\over {\pad t}} 
+ { {\left (
\eta_{tt} 
+ \epsilon_r r_0 {\ds\frac{\pad \eta_{tt}}{\pad r}} 
+ h_{tt} 
+ \epsilon_r r_0 {\ds\frac{\pad h_{tt}}{\pad r}} 
+ \epsilon_t t_0 {\ds\frac{\pad h_{tt}}{\pad t}}  
\right )} 
\over {\pad r}} 
\right ] 
\frac{dr}{ds}\frac{dt}{ds}  
\right \}
\]
\[
+ {1 \over 2} 
\left (
h^{rt} 
+ \epsilon_r r_0 {\ds\frac{\pad h^{rt}}{\pad r}} 
+ \epsilon_t t_0 {\ds\frac{\pad h^{rt}}{\pad t}}  
\right ) \times
\]
\[
\left \{
\pad \left  [
{{2 \left (
h_{rt} 
+ \epsilon_r r_0 {\ds\frac{\pad h_{rt}}{\pad r}} 
+ \epsilon_t t_0 {\ds\frac{\pad h_{rt}}{\pad t}}  
\right )} 
\over {\pad r}} 
+ {{\left (
h_{rr} 
+ \epsilon_r r_0 {\ds\frac{\pad h_{rr}}{\pad r}} 
+ \epsilon_t t_0 {\ds\frac{\pad h_{rr}}{\pad t}}  
\right )} 
\over {\pad t}} 
\right ] 
\frac{dt}{ds}\frac{dr}{ds}  
\right .
\]
\[
+ \pad \left [
{{2 \left (
h_{rt} 
+ \epsilon_r r_0 {\ds\frac{\pad h_{rt}}{\pad r}} 
+ \epsilon_t t_0 {\ds\frac{\pad h_{rt}}{\pad t}}  
\right )} 
\over {\pad t}} 
+ 
{{ \left (
\eta_{tt} 
+ \epsilon_r r_0 {\ds\frac{\pad \eta_{tt}}{\pad r}} 
+ h_{tt} 
+ \epsilon_r r_0 {\ds\frac{\pad h_{tt}}{\pad r}} 
+ \epsilon_t t_0 {\ds\frac{\pad h_{tt}}{\pad t}}  
\right )} 
\over {\pad r}} 
\right ]
\frac{dt}{ds}\frac{dt}{ds} 
\]
\be
\left .
+ \pad\left [
{ {3 \left (
h_{tt} 
+ \epsilon_r r_0 {\ds\frac{\pad h_{tt}}{\pad r}} 
+ \epsilon_t t_0 {\ds\frac{\pad h_{tt}}{\pad t}}  
\right )} 
\over {\pad t}} 
\right ]
\frac{dt}{ds}\frac{dt}{ds} \right \} = 0
\label{eq:dr2ds2}
\ee
}

The quantity {\small 
${d^2r}/{dt^2}$} is deduced as ratio from eq. (\ref{eq:dr2ds2}), 
and from the same eq. in where {\small $\mu = t$}.
Eq. (\ref{eq:dr2ds2}) at
lowest order ({\small $h = h' = h'' = \eta'' = 0 $}) 
expresses the known 
motion of a particle in a unperturbed metric. Otherwise,  
the particle crosses the perturbed metric and thus 
its motion is influenced by the emitted radiation. 
The velocities {\small $\dot r$} and {\small $\dot t$} 
are derived from the Lagrangian.   
The field contribution at 1st-order is: 

{\small 
\be
{\cal L}_1  = 
\!\left (\!1\!-\!{2M\over r}\!\right )\!H_0Y{\dot t} ^2 
+ \!\left (\!1\!-\!{2M\over r}\!\right )^{-1}H_2Y{\dot r} ^2
+ r^2 KY {\dot \theta}^2 
+ r^2 \sin^2\theta KY  {\dot \phi}^2 
+ 2 H_1 Y {\dot t} {\dot r} 
\label{eq:Lagrf} 
\ee
}
where the dot indicates {\small $s$} derivatives. 
Differentiating {\small ${\cal L}_1$} 
(omitting the index {\small $l$}) and eliminating angular 
dependence, the canonical quantities are:

{\small 
\be
{\dot p}_t = \frac{\pad {\cal L}_1}{\pad t} =
\!\left (\!1\!-\!{2M\over r}\!\right )\!\frac{\pad H_0}{\pad t}Y{\dot t}^2 
+ \!\left (\!1\!-\!{2M\over r}\!\right )^{-1}
\frac{\pad H_2}{\pad t}Y{\dot r}^2
+ 
2 \frac{\pad H_1}{\pad t} Y {\dot t}
{\dot r}
\label{eq:dpt}
\ee

\be
p_t = \frac{\pad {\cal L}_1}{\pad  {\dot t}} = 
2 \!\left (\!1\!-\!{2M\over r}\!\right )\!H_0Y{\dot t}
+ 2 H_1 Y{\dot r}
\ee


\be
{\dot p}_r\! = \! - \! 
\frac{\pad {\cal L}_1}{\pad r}\! = \!
 - \! \left [\!\frac{2M}{r^2} H_0 
\! + \!\left (\!1\!-\!{2M\over r}\!\right )\!\frac{\pad H_0}{\pad r} \!
\right ]\!
Y{\dot t}^2 
\!+\! \left [\!
\frac{2M}{(r\!-\!2M)^2} H_2
\!-\! \left (\!1\!-\!{2M\over r}\!\right )^{-1}\frac{\pad H_2}{\pad r}\!
\right ]\! 
Y{\dot r}^2
\! - \! 2 \frac{\pad H_1}{\pad r}Y{\dot t} {\dot r} 
\ee

\be
p_r = - \frac{\pad {\cal L}_1}{\pad {\dot r}} = 
- 2\!\left (\!1\!-\!{2M\over r}\!\right )^{-1}H_2Y{\dot r} - 2 H_1{\dot t} Y
\ee

}

The Hamiltonian 
is equal to the Lagrangian due to the absence of the   
potential. 
The Lagrangian is time independent for a conservative system: 
the power of the emitted radiation $P_{gw}$ is  
added to (\ref{eq:dpt}) 
{\small $
{\dot p}_{t, total} = {\dot p}_{t} + P_{gw} = 0$}. 
The integration in time leads to 
a constant of energy dimensions from which 
{\small $\dot t$} is derived:

{\small 
\be
p_{t, total} = 
\int 
\!\left (\!1\!-\!{2M\over r}\!\right )\!\frac{\pad H_0}{\pad t}Y{\dot t}^2 
+ \!\left (\!1\!-\!{2M\over r}\!\right )^{-1}
\frac{\pad H_2}{\pad t}Y{\dot r}^2
+ 
2 \frac{\pad H_1}{\pad t} Y {\dot t}
{\dot r}dt + \int P_{gw} dt
\label{eq:pt2}
\ee
}

The Lagrangian is unitary for timelike geodesics and  
substituting (\ref{eq:pt2}) in (\ref{eq:Lagrf}) 
for 
{\small $\dot \theta = \dot \phi = 0$}, 
{\small ${\dot t}(r)$} and 
{\small $\dot r$} are calculated.
The initial conditions at infinity, {\small ${\dot r} = 0$} and 
{\small $E_{gw} = 0 $}, fix the constant $E$.
The zeroth order terms must be added to the 1st-order ones.  

\subsection{Singularities}

The energy going into the black hole is 
{\small $E_{in} \simeq 0.3927m c^2$} 
while the 
 energy radiated away is
{\small $E_{out} \simeq 0.0104 {\mu^2c^2}/{M}$} 
where the reduced mass 
{\small $\mu \simeq m$}. The 
relation between {\small $E$} and {\small $\psi$} or:

{\small
\be
T^{\mu\nu} = \frac{1}{32\pi} h^{*}_{\rho\sigma ;\mu}
                             h^{\rho\sigma}_{;\nu}
\ee
}
should bring to the normalization of divergent expressions. 
Alternatively a finite size mass, e.g. dust, could be considered. 

\section{Conclusions}

The approach for identification of radiation reaction of masses falling  
into Schwarz\-schild black holes via 1st-order polar perturbations 
has been shown and 
the equations of motion in symbolic form have been found. 
Radiation reaction is a fundamental concept 
in bodies motion theory, but also has relevant 
implications on detector's templates since the capture of 
stars by black holes is   
a source of gravitational waves.

\section*{Acknowledgements} 

Discussions with V. Pierro and I. Pinto (Salerno), 
G. Sch\"{a}fer (Jena) are acknowledged. Encouragement from S. Chandrasekhar  
was the warmest gift. Financial support for participating at this 
conference was received from 
the European Space Research \& Technology Centre, Noordwijk.

\end{document}